\documentstyle[prb,aps,epsf,twocolumn]{revtex}
\begin{document}
\draft
\wideabs{
\title{Electronic structure of the strongly hybridized
ferromagnet CeFe$_2$}
\author{T. Konishi\cite{Chiba-U}, K. Morikawa, K. Kobayashi, T. Mizokawa and A. Fujimori}
\address{Department of Physics and Department of Complexity Science
and Engineering,\\ University of Tokyo, Bunkyo-ku, Tokyo 113-0033, Japan}
\author{K. Mamiya}
\address{Hiroshima Synchrotron Radiation Research Center,
Hiroshima University,\\ Higashi-Hiroshima 739-8526, Japan}
\author{F. Iga}
\address{Department of Quantum Matter, ADSM, Hiroshima University,\\
Higashi-Hiroshima 739-8526, Japan}
\author{H. Kawanaka}
\address{Electrotechnical Laboratory, Tsukuba, Ibaraki 305-8568, Japan}
\author{Y. Nishihara}
\address{Faculty of Science, Ibaraki University, Mito, Ibaraki 310-8512,
Japan}
\author{A. Delin}
\address{Fritz-Haber-Institut der Max-Planck-Gesellschaft  \\
 Faradayweg 4-6, D-14 195 Berlin-Dahlem, Germany}
\author{O. Eriksson}
\address{Department of Physics, Uppsala University \\
 P.~O. Box 530, S-751 21 Uppsala, Sweden}
\date{\today}
\maketitle
\begin{abstract}
We report on results from  high-energy spectroscopic measurements on
CeFe$_2$, a system of particular interest due to its
anomalous ferromagnetism with an unusually low Curie temperature and
small magnetization compared to the other rare earth-iron Laves phase
compounds.
Our experimental
results, obtained using core-level and valence-band
photoemission, inverse photoemission and soft x-ray absorption
techniques,
indicate very strong hybridization of the Ce 4$f$ states with the
delocalized band states, mainly the Fe 3$d$ states.
In the interpretation and analysis of our measured spectra, we have
made use of
two different theoretical approaches: The first one is
based on the Anderson impurity model, with surface contributions
explicitly taken into account.
The second method
consists of band-structure calculations
for bulk CeFe$_2$.
The analysis based on the Anderson impurity model
gives calculated spectra in good agreement
with the
whole range of measured spectra, and
reveals that the Ce 4$f$ - Fe 3$d$ hybridization is considerably
reduced at the surface, resulting in even stronger hybridization in the
bulk than previously thought.
The band-structure calculations are
{\it ab initio} full-potential linear muffin-tin orbital
calculations within the
local-spin-density approximation
of the density functional.
The Ce 4$f$ electrons
were treated as itinerant band electrons.
Interestingly,
the Ce 4$f$ partial density of states obtained
from the
band-structure calculations
also agree well with the experimental spectra
concerning both the 4$f$ peak position and the 4$f$ bandwidth, if the
surface effects are properly taken into account.
In addition, results, notably the partial
spin magnetic moments, from the
band-structure calculations are discussed in some detail and compared to
experimental findings and earlier calculations.
\end{abstract}
\pacs{PACS numbers: 71.27.+a, 75.75.Bb, 75.30.Mb, 79.60.Bm}
}
\narrowtext
\section{Introduction}
The 4$f$ states of rare-earth elements in solids usually retain
free-ionic properties with a well-defined integer occupation number.
However, there are also rare-earth compounds where the
hybridization of the 4$f$ states with extended band states is
important, in which case they may exhibit properties
usually only found in actinide systems. In such systems, many unusual
phenomena are
typically observed, like for instance
anomalously low saturation magnetization and
Curie temperature $T_C$ (e.g., CeFe$_2$),
intermediate valence (e.g, SmS), heavy fermion
behavior (e.g., YbBiPt), or
non-Fermi-liquid behavior (e.g., CeCu$_{6-x}$Au$_x$).
Even more surprisingly, simultaneous magnetic ordering and
superconductivity has been observed (e.g., CeCu$_2$Si$_2$).
The superconductivity is unconventional, i.e., 
the order parameter suggests a d-wave superconducting state, as opposed
to the conventional s-wave state. 
Finding a proper theoretical description
of the 4$f$ states in these compounds remains one of the major problems
in condensed matter physics.\cite{amato97}

As for the
photoemission spectroscopy (PES) studies of Ce compounds, it is widely
believed that the spectra are well described by the single-impurity
Anderson model (SIAM).~\cite{Allen}
Recently, however, it has been argued that systems, in which
the Ce 4$f$ states hybridize strongly with the other valence
electrons,
calculations based on density functional theory (DFT) may
give an equally good, or even better description of the
photo-emission spectra than the SIAM analysis, provided that surface effects are
properly taken into account in the analysis.~\cite{Weschke}
However, one should bear in mind that calculations based on DFT
are not strictly applicable for excited state properties,
instead the ground state properties, such as the magnetic moments of the ground state,
which are typically the focus of these calculations.
Nevertheless, the electronic structure given from
such calculations are often compared with photoemission data
and good agreement between
experiment and calculations is frequently observed.
In the limit of complete screening of the excited state, one
would expect ground-state density functional
calculations to be able to describe the spectra well.

CeFe$_2$ is thought to belong to a class of strongly hybridized systems.
This compound shows ferromagnetism below $T_C$ = 230 K with a
saturation magnetization of 2.30 $\mu_B$/f.u. Above
$T_C$, the magnetic susceptibility follows the Curie-Weiss law with
an
effective moment of 7.4 $\mu_B$/f.u.~\cite{Deportes}
If one compares CeFe$_2$ with the other $R$Fe$_2$ compounds
($R$: rare earth elements), a number of anomalies in its
physical properties can be observed.
The lattice constant is much smaller than
an interpolation using the lattice constants of the other $R$Fe$_2$
systems would suggest.
Its Curie temperature
is anomalously low: the other $R$Fe$_2$ compounds
have Curie temperatures ranging from 596 K to 796 K.~\cite{Buschow}
The saturation magnetization is unusually low compared to the other $R$Fe$_2$ compounds (2.93 and
2.90$\mu_B$/f.u. for LuFe$_2$ and YFe$_2$, respectively~\cite{Buschow}).
Moreover, even if only a small fraction of the Fe atoms
are substituted for Al, the ferromagnetic ordering is destroyed,
and the system becomes anti-ferromagnetic.~\cite{Nishihara}
In fact, even in pure CeFe$_2$, recent neutron scattering experiments have
revealed strong competition between the ferromagnetic ground state and
an antiferromagnetic ground state.\cite{paolasini98}
Together,
these facts suggest that the Ce 4$f$ states in CeFe$_2$
hybridize strongly with the other
valence electrons, notably the Fe $3d$ valence states.
This hypothesis is further supported by the
X-ray absorption (XAS) experiments by Croft et al.~\cite{Croft}

In this paper, we present high-energy spectroscopic results on CeFe$_2$
including core-level x-ray photoemission (XPS), XAS, Ce 3$d$-4$f$ and
4$d$-4$f$ resonant PES, bremsstrahlung isochromat (BIS) and
high-resolution ultraviolet photoemission spectroscopy (UPS) in order
to elucidate the electronic structure of this system. In the case of Ce
compounds with strongly hybridized 4$f$ states, it has been pointed out
that surface effects are extremely important in the interpretation of the
spectra.~\cite{Laubschat,Weschke2,Laubschat2} Therefore, we have
attempted to differentiate the electronic structure of bulk and that of
surface for CeFe$_2$ in the analysis of the spectra.
As will be further elaborated on in section~\ref{bs_results}
of this paper,
electronic-structure calculations with the Ce
4$f$ states treated as valence states give a good description of
the magnetism in CeFe$_2$.~\cite{eriksson88,trygg94}
It is of course highly interesting to assess the applicability of
the same theory in describing also the
photoemission spectra of CeFe$_2$, even though as noted these calculations are
not strictly applicable for excited state properties.
Very recently, Sekiyama {\it et al.}~\cite{sekiyama} reported a high-resolution
3$d$-4$f$ resonant photoemission study of the strongly hybridized
system CeRu$_2$ and found that the Ce 4$f$ spectra can be explained by
band theory.
In the following, we first attempt
to describe the spectra in the framework of the SIAM and obtain a set
of SIAM parameters. In the analysis, surface effects on the spectra are
explicitly taken into account. Next, the bulk component of the
valence-band spectra is compared with the density of
states calculated using density functional theory. All DFT results
presented here have been calculated within the local spin-density
approximation (LSDA). The use of
more recently developed generalized gradient functionals would not,
however, alter any of our conclusions.

\section{Methods}
\subsection{Experiment}
Polycrystalline samples of CeFe$_2$ were prepared by arc-melting the
pure constituent materials. Subsequently, the samples were annealed at
750 $^{\circ} $C for a week to obtain single phase samples. Magnetization
measurements yielded the same $T_C$ as in the literature. The XPS spectra
were taken with Mg $K{\alpha}$ radiation ($h\nu$ = 1253.6 eV) using a
double-pass cylindrical-mirror analyzer, and
the BIS spectra were obtained using
a Pierce-type electron-gun and a quartz crystal monochromator which was
set at $h\nu$ = 1486.6 eV. The Ce 4$d$-4$f$ resonant PES measurements
were done at beam-line BL-2 of SOR-RING, Institute for Solid State
Physics, University of Tokyo. The Ce 3$d$-4$f$ resonant PES and Ce 3$d$
XAS data were taken at beam-line BL-2B of Photon Factory, High Energy
Accelerator Research Organization. Photoelectrons were collected using
a double-pass cylindrical-mirror analyzer in the resonant PES
measurements. The XAS spectra were obtained by measuring the total
electron yield using an electron multiplier placed near the sample. All
measurements were done in the range 50-80 K, i.e., below the Curie
temperature. In the case of the
Ce 4$d$-4$f$ resonant PES, additional measurements at room
temperature, i.e, above the Curie temperature, were performed.
The total energy resolution was $\sim$1.0 eV for XPS and
BIS, $\sim$0.5 eV for Ce 4$d$-4$f$ resonant PES, $\sim$0.5 eV for XAS
and $\sim$1.0 eV for Ce 3$d$-4$f$ resonant PES. The high-resolution UPS
measurements were done around 17 K using a hemispherical analyzer and
the He~{\small I} ($h\nu$ = 21.2 eV) and He~{\small II} ($h\nu$ = 40.8
eV) resonance lines. The energy resolution was $\sim$25 meV for both
photon energies. The binding energies were calibrated using Au
evaporated on the samples. For XAS and Ce 3$d$-4$f$ resonant PES, the
photon energies were calibrated using the Cu 2$p$ edge of Cu metal and
the Co 2$p$ peak of LaCoO$_3$.
Clean surfaces were obtained by scraping
the sample repeatedly, while maintaining the sample under ultra-high vacuum,
with a diamond file prior to each measurement.
Cleanliness of the surfaces was
checked by the absence of O 1$s$ and C 1$s$ XPS signals from
contaminants in the case of the XPS, XAS, BIS and Ce 3$d$-4$f$ resonant
PES measurements.
In the case of the high-resolution UPS and Ce
4$d$-4$f$ resonant PES measurements, cleanliness was checked by the
absence of a O 2$p$ feature which appears around 6 eV below the
Fermi level ($E_F$).

\subsection{Single-impurity Anderson Model }
The SIAM calculations were made based on the
variational 1/$N_f$-expansion method developed by Gunnarsson and
Sch\"onhammer.~\cite{GS} Here, we performed the calculations
to the lowest order in 1/$N_f$, where $N_f$ is the degeneracy of the Ce
4$f$ level and was taken to be 14. The $f^2$ configuration was also
included in the calculation. The energy dependence of the hybridization
matrix elements was taken from the off-resonant spectra, which
approximately represent the Fe 3$d$ partial density of states. The
configuration dependence of the hybridization strength was also taken
into account, and was chosen to be the same as that obtained for
$\alpha$-Ce by Gunnarsson and Jepsen.~\cite{GJ} In the calculations, we
divided the band continuum into discrete levels following Kotani {\it
et al.}~\cite{Kotani} We further assumed that each spectrum was a
superposition of two components which represent bulk and surface
spectra. The weight of each component was treated as fitting parameters
within a range consistent with the universal curve for the mean free
path of photoelectrons.~\cite{Shirley}

\subsection{Band-Structure Calculation}
In the band-structure calculations presented here, we have
used the
full-potential linear muffin-tin orbital method (FP-LMTO).~\cite{wills}
In this method, the Kohn-Sham equations~\cite{dft}
are solved for a general potential without any
shape approximation.
The local (spin)
density approximation (LSDA) in the Hedin-Lundqvist
parameterization~\cite{hedin71} was used for the density functional.

In the FP-LMTO method, space is divided into non-overlapping
spheres, so-called called muffin-tin spheres,~\cite{lmto}
surrounding each atomic site, and an interstitial region.
The basis functions used are energy independent Bloch functions, whose
construction is different in the spheres and in the
interstitial.

A basis function in the interstitial is defined by the
Bloch function of solutions to the spherical Helmholtz equation with
nonzero
kinetic energy $\kappa^2$, or a linear combination of such solutions for
different kinetic energies.
The Fourier representation of this basis function
is taken from the Fourier series of a
function matching the basis in the interstitial region but not inside
the spheres, a so-called pseudo-wave function, whose
exact shape inside the muffin-tin sphere is of
no importance for the final solution as long as it is continuous and
differentiable at the sphere boundary and matches the true basis
function in the interstitial.

Inside the spheres, where the charge density varies rapidly,
the basis functions are Bloch functions of numerical radial functions times
spherical harmonics.
The radial part of a basis function is constructed
from
the numerical solutions $\phi_L(E_\nu,r)$
of the radial Schr\"odinger equation in a spherical potential
at
the fixed energy $E_\nu$,
and their energy derivatives $\dot{\phi}_L(E_\nu,r)$.
Here, the index $L$ stands for a collection of quantum numbers: the
principal quantum number $n$, the orbital quantum number $l$,
the magnetic quantum number $m$,
and the kinetic energy $\kappa^2$.

The treatment of the entire basis set within one single energy panel
allows all states, including the
semi-core states, to hybridize fully with each other.
Our method is linear, i.e., the basis functions are constructed by
expanding around fixed energies $E_\nu$.
The expressions for the crystal wave functions in the muffin-tin spheres
are
matched to the interstitial crystal wave function
at the sphere boundaries so that the
total crystal wave function becomes continuous and differentiable in all
space.
In the present calculation,
the expansion in
spherical harmonics was taken up to $l = 6$.
For Ce, the
$6s$, $5p$, $6p$, $5d$, and $4f$ orbitals were included in the basis set,
with  $5p$ as semi core.
For Fe, we included the $4s$,  $4p$ and  $3d$ orbitals, i.e., no
semi-core state was used for Fe.
Four $\kappa^2$-values were used in the calculation:
-0.6 Ry and -0.1 Ry for the
valence states, and -1.5 Ry and -1.0 Ry
for the semi-core Ce $5p$ states, all with
respect to the muffin-tin zero.

Reciprocal space was sampled with what would correspond to 1331
{\bf k}-points in the full Brillouin zone (BZ)
using special {\bf k}-point sampling
methods.~\cite{spec-k}
The non-overlapping muffin-tin spheres were
chosen as 21\% and 17\% of the unit-cell lattice constant
for Ce and Fe, respectively.
With this choice,
36\% or the unit cell volume is
in the interstitial region and the
closest muffin-tin spheres are 3\% from touching.

The experimental lattice constant was used in the
calculations.
Furthermore, the calculations were spin-polarized but
the spin-orbit interaction was not included.
This latter approximation will be commented on further in
conjunction with presenting and discussing the results from the
band-structure calculation.

\begin{figure}
  \centering\epsfxsize=0.5\textwidth \epsffile{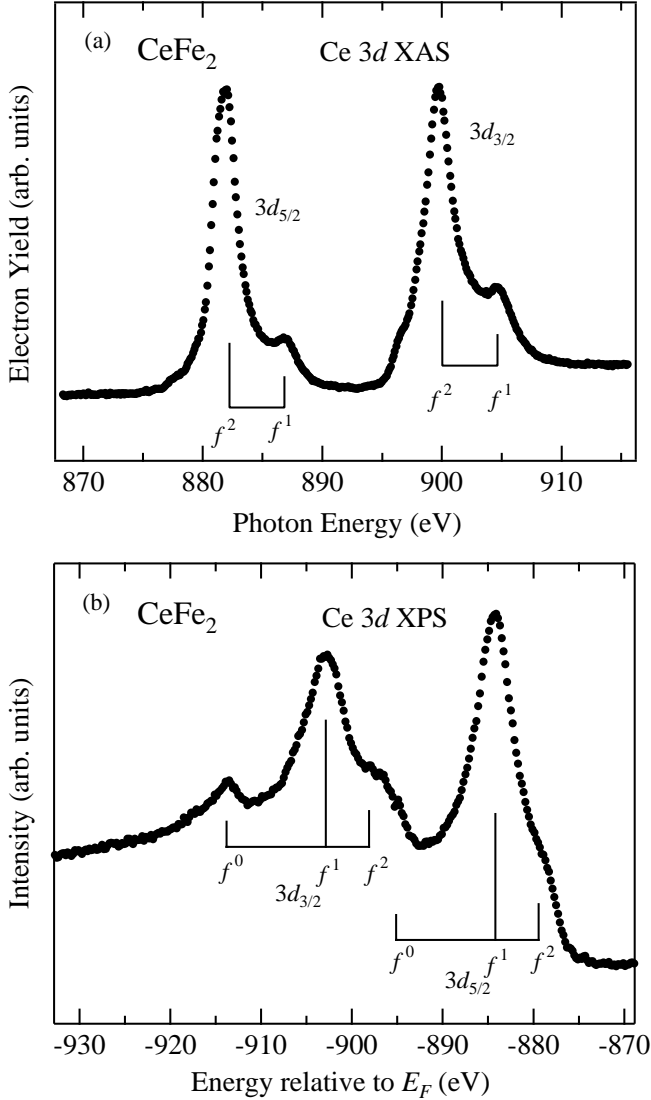}
\caption{
Core-level spectra of CeFe$_2$. (a) Ce 3$d$ core-level XAS spectra.
(b) Ce 3$d$ XPS spectra taken at $h\nu$ = 1253.6 eV.
}
\label{core.fig}
\end{figure}
\section{results}
\subsection{Experiment}
Fig.~\ref{core.fig} shows the Ce 3$d$ core-level XPS and XAS spectra. The
XPS line-shape is a typical one for a strongly hybridized Ce compound,
consisting of three peaks which correspond to the 3$d^9$4$f^0$,
3$d^9$4$f^1$ and 3$d^9$4$f^2$ final states in each of the $j$=3/2 and
5/2 spin-orbit components~\cite{baer78,fuggle80}.
In the XAS spectrum, the main peaks are due
to  the 3$d^9$4$f^2$ final-state multiplet and the satellite structures
$\sim$5 eV above the main peaks are due to the 3$d^9$4$f^1$ final
states. The rather distinct 4$f^0$ peaks in the XPS spectrum and the
4$f^1$ structures in the XAS spectra, together with the obscured
3$d^9$4$f^2$ final-state multiplet structures of the main XAS peaks,
indicate strong hybridization of the 4$f$ states with the valence band
in this system. The XPS spectrum reflects the surface
electronic structure because of the rather low kinetic energies of
photoelectrons from the Ce 3$d$ core level. A detailed analysis of this is given
below.

\begin{figure}
  \centering\epsfxsize=0.5\textwidth \epsffile{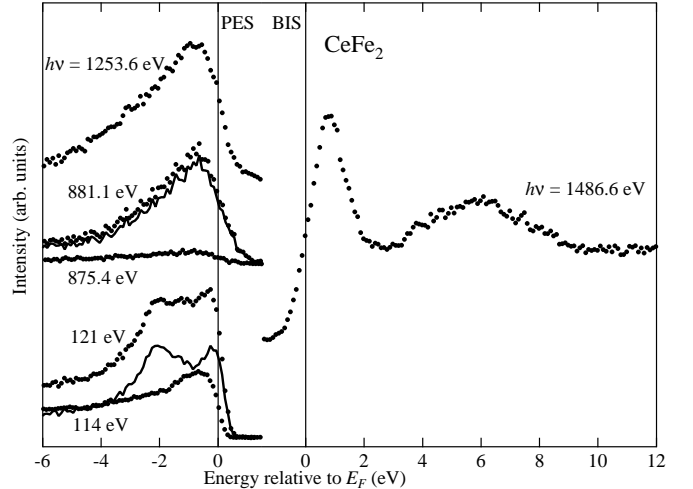}
\caption{
Valence-band PES and BIS spectra of CeFe$_2$. $h\nu =$ 121 and 114 eV
(881 and 875 eV) correspond to Ce 4$d$-4$f$ (3$d$-4$f$) on- and
off-resonance, respectively. Solid curves show the difference spectra,
which represent the Ce 4$f$ component.
}
\label{pes_bis.fig}
\end{figure}
The results of valence-band PES and BIS are shown in Fig.~\ref{pes_bis.fig}.
The on- and off-resonance occurs, respectively, at $h\nu$ = 121 and 114
eV in the Ce 4$d$-4$f$ resonant PES and at $h\nu$ = 881.1 and 875.4 eV
in the Ce 3$d$-4$f$ resonant PES. Identical spectra in the present
resolution have been obtained for Ce 4$d$-4$f$ resonant PES at room temperature,
which is above $T_C$ (not shown). We have obtained the Ce 4$f$ spectra
by subtracting the off-resonance spectra from the on-resonance spectra
as shown by solid curves in Fig.~\ref{pes_bis.fig}. As seen from this figure,
there is a large difference between the Ce 4$f$ spectra obtained
from the Ce 4$d$-4$f$ and Ce 3$d$-4$f$ resonant PES. While the former
has a double-peak structure in the vicinity of -2 eV and near $E_F$, the latter
is dominated by a single peak near $E_F$, implying stronger
hybridization in the latter. This can be attributed to the difference
in the surface sensitivity of the two spectra due to the different
kinetic energies of photoelectrons. This also indicates that the Ce
valency at the surface is closer to trivalent than it is in the bulk.

In Fig.~\ref{pes_bis.fig} we also show the XPS spectrum of the valence band
taken with Mg $K\alpha$ radiation. Owing to the higher kinetic energies
of photoelectrons, this spectrum is considered to be more bulk sensitive
than the above PES spectra.
Considering the photoionization cross-sections,~\cite{Yeh} the valence-band XPS spectrum
should mainly reflect
the Fe 3$d$ partial density of states (DOS)
with significant contributions
from Ce 4$f$ and Ce 5$d$. As seen in the figure, the XPS spectrum shows
a line-shape similar to the off-resonance spectra of Ce 4$d$-4$f$ and Ce
3$d$-4$f$ resonant PES.

It is expected, that the BIS spectrum should also reflect the bulk
electronic structure rather
well. There is a peak near $E_F$ and a broader feature at $\sim$6 eV.
They originate mainly from the  Ce 4$f$ states, although there are
contributions from the Ce 5$d$ and Fe 3$d$ states, too. The peak near
$E_F$ and the structure around $\sim$6 eV correspond to the 4$f^1$ and
4$f^2$ final states, respectively.~\cite{Allen} The broad line-shape of
the structure at $\sim$6 eV is due to the 4$f^2$ final-state
multiplet.~\cite{Allen} The strong intensity of the peak near $E_F$
again indicates strong hybridization of the Ce 4$f$ states with the
valence states.

\begin{figure}
  \centering\epsfxsize=0.5\textwidth \epsffile{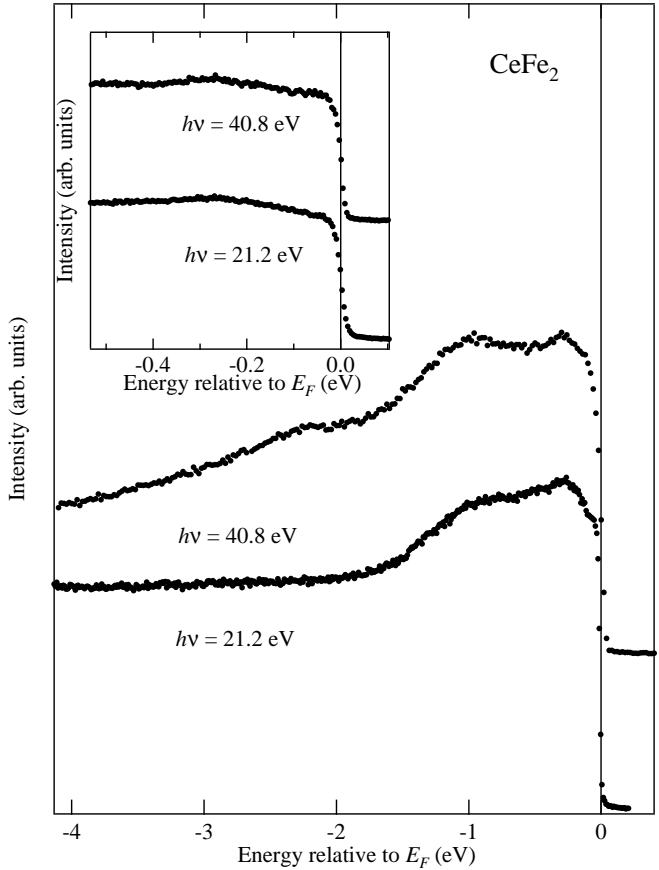}
\caption{
High-resolution UPS spectra of CeFe$_2$. Inset shows an enlarged view
near $E_F$.
}
\label{ups.fig}
\end{figure}
Fig.~\ref{ups.fig} shows high-resolution UPS data. In this photon energy
range, the cross sections of the Fe 3$d$, Ce 4$f$ and Ce 5$d$ states
varies rapidly with photon energy.~\cite{Yeh} The relative cross-section
of Ce 4$f$ to the other orbitals increases when going from $h\nu$ =21.2
to 40.8 eV while that of Ce 5$d$ rapidly decreases. Therefore the
structure at -(2 $\sim$ 3) eV which appears only in the $h\nu$ = 40.8 eV
spectrum, originates from  the Ce 4$f$ states, and corresponds to one of
the double peaks in the Ce 4$f$ spectrum obtained by the Ce 4$d$-4$f$
resonant PES. This observation is also consistent with the fact that the
40.8 eV spectrum is surface sensitive,
according to the ``universal curve'' of the mean free path of
photoelectrons.~\cite{Shirley} In the near $E_F$ region, structures
just below $E_F$ and at $\sim$ -0.3 eV are somewhat enhanced in the 40.8
eV spectrum.
These structures originate from the Ce 4$f$ states
and correspond to the tail of the Kondo resonance (possibly with
unresolved fine structures due to crystal-field splitting) and the
spin-orbit side band, respectively.
These structures are also expected
to be dominated by surface contributions.
\subsection{Single-Impurity Anderson Model}
\begin{figure}
  \centering\epsfxsize=0.5\textwidth \epsffile{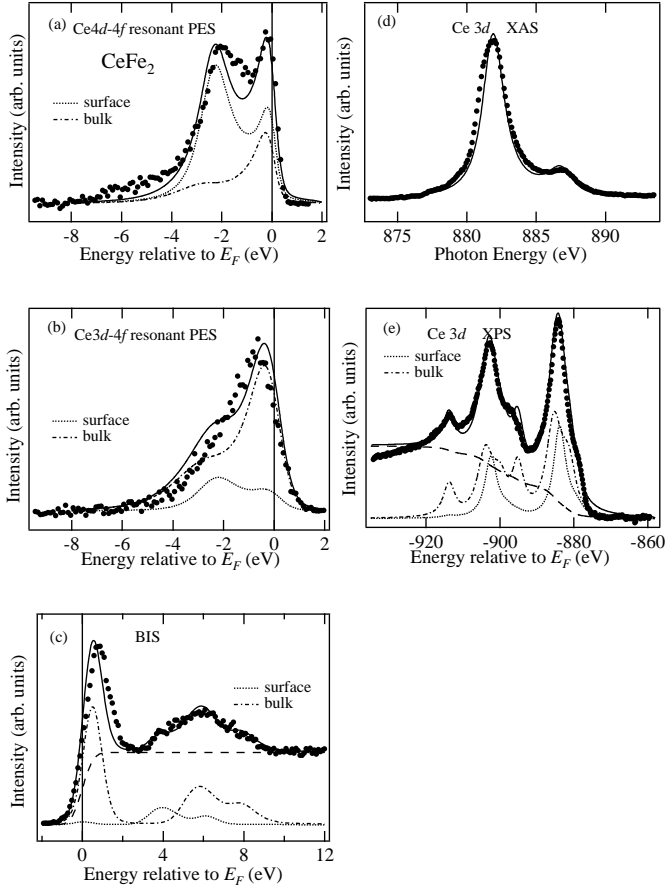}
\caption{
Comparison of the single-impurity Anderson model calculation with the
experimental spectra of CeFe$_2$. (a) Ce 4$f$ spectrum obtained from Ce
4$d$-4$f$ resonant PES. (b) Ce 4$f$ spectrum obtained from Ce 3$d$-4$f$
resonant PES. (c) BIS spectrum. (d) Ce 3$d$ core-level XAS spectrum. (e)
Ce 3$d$ core-level XPS spectrum. In each panel, dots show experimental
spectrum, solid curve shows the calculated spectrum, and dotted and
dash-dotted curves show the calculated surface and bulk components,
respectively.
}
\label{siam.fig}
\end{figure}

The SIAM parameters obtained in our calculation are
listed in Table~\ref{table1}.
Here, $\epsilon_f$ is the position of the bare 4$f$ level
(4$f^1$ $\to$ 4$f^0$ ionization level) relative to $E_F$,
$U_{\it ff}$ is the 4$f$-4$f$
on-site Coulomb energy, $U_{\it fc}$ is the 4$f$-core-level Coulomb energy
and $V$ is the Ce 4$f$-valence-band hybridization strength, in accordance
with the definitions in Ref.~\onlinecite{Kotani}.
Using those parameters, the $4f$ occupation number $n_f$ has been
calculated and listed in the last  column of Table~\ref{table1}.
Since we have fitted many different types of spectra using the SIAM
shown in Fig.~\ref{siam.fig}, many constraints have lead to a rather
unique set of SIAM parameters.

In Fig.~\ref{siam.fig}, comparison is made between the SIAM calculations and the
experimental spectra.
As seen, we obtain good overall agreement with all
experimental spectra.
The main discrepancy between the SIAM results and experimental
spectra is found in the BIS spectrum (Fig.~\ref{siam.fig}c),  where
the position of the calculated $f^1$ peak is about
0.5 eV lower than in the experimental spectrum.
Noticeable from this figure is also the large difference between the bulk and surface spectra
obtained through the SIAM analysis.
For instance, in the Ce $4f$ spectrum obtained from Ce $4d$-$4f$
resonant PES (Fig.~\ref{siam.fig}b), which is basically a double-peak structure,
the relative strength of the two peaks is very different between
the bulk spectrum and the
surface spectrum.
Similar differences in directly measured bulk- and surface spectra
have previously been reported for Ce-metal,\cite{Weschke2} and can be explained
as due to larger hybridization in the bulk.\cite{GS}

Table~\ref{table1} shows that, apparently, the Ce atoms belonging to the
surface are as good as completely trivalent,
with  $n_f \simeq 1.0$, whereas in the bulk, the
4$f$ states are strongly hybridized, having the
significantly lower occupation of 0.78.
\begin{table}
\caption{
SIAM parameters for CeFe$_2$.
$\epsilon_f$, $U_{\it ff}$, $V$, and $U_{\it fc}$ are
given in units of eV.
}
\label{table1}
\begin{tabular}{lccccc}
&$\epsilon_f$&$U_{\it ff}$&$V$ &$U_{\it fc}$&$n_f$\\
\tableline
surface&-1.8&6.4&0.23&9.7&1.0\\
bulk&-0.8&6.4&0.41&9.7&0.78\\
\end{tabular}
\end{table}
However, there are noticeable
amplitudes of the $f^0$ and $f^2$ configurations also at the surface,
indicating that also here, some hybridization between
the $4f$ and valence states is taking place.
\subsection{Band-Structure Calculation}
\label{bs_results}
%
%
Experimentally, the partial moments in CeFe$_2$ have been studied
using several different experimental methods:
polarized neutrons, \cite{kennedy93}
Compton scattering\cite{cooper96} and, very recently,
x-ray magnetic circular dichroism (XMCD).\cite{delobbe98}
In all experiments, an anti-parallel coupling of the
Ce and Fe moments is found. This coupling is also reproduced in our
calculation, as well as in earlier calculations.\cite{eriksson88,trygg94}
As is well known,\cite{eriksson88} this
anti-parallel coupling of the moments is a strong indication that
the Ce $4f$ states in CeFe$_2$ are delocalized. This can easily be understood from the
following argumentation. If the $4f$ electrons
are localized, the $4f$ spin moment would be dictated by the polarization of the
$spd$-electrons of the Ce atom, that via hybridization effects are known to be anti-parallel
to the $3d$ moment of the Fe atom. Hence the {\it spin} moments of the Ce atom and the Fe atom
are always anti-parallel, both in the localized and delocalized case.
For localized $4f$ electrons, the $4f$ spin moment is
accompanied by an orbital moment (larger than the spin moment)
that (via Hund's third rule) is anti-parallel to the Ce spin
moment. Hence, for localized $4f$ electrons the total (spin+orbital)
Ce-Fe coupling is ferromagnetic,
whereas if the Ce $4f$
orbital moment is quenched, due to band formation, the coupling is anti-parallel.

\begin{figure}
  \centering\epsfxsize=0.45\textwidth \epsffile{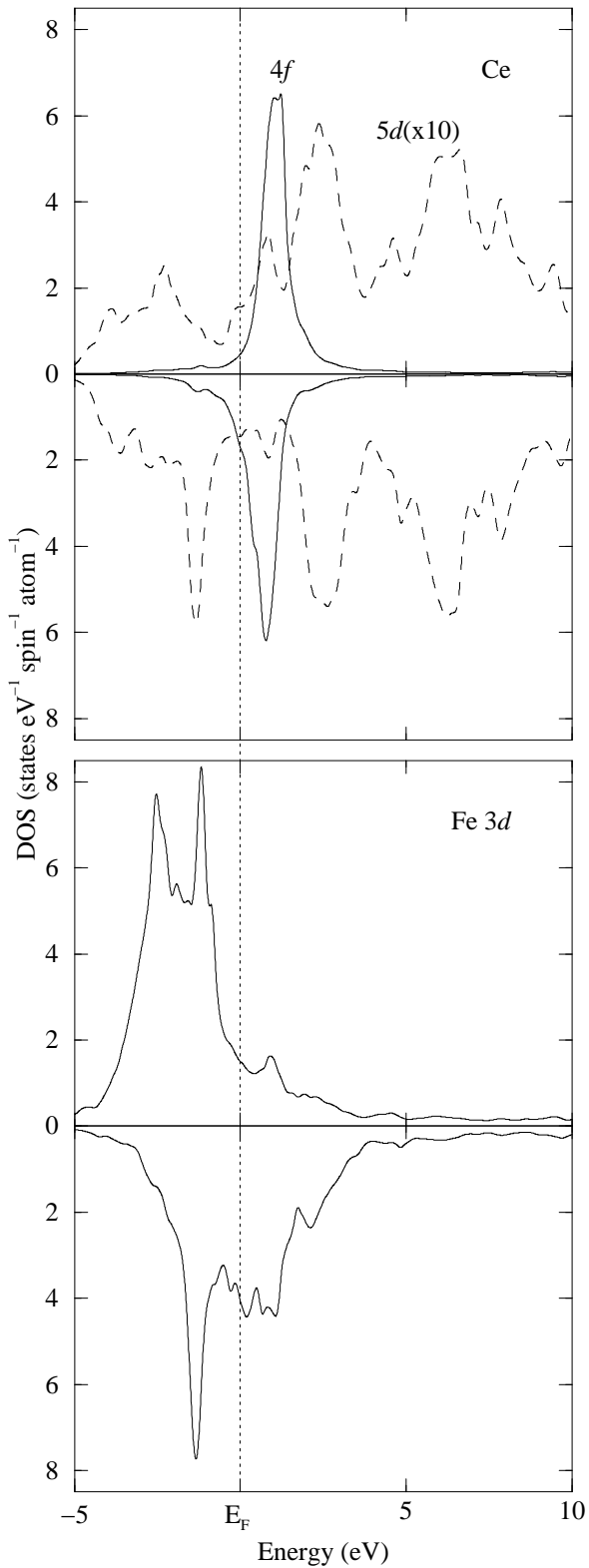}
 \caption
 {
 \label{dos.fig}
 Spin-resolved partial DOS for the
 Ce $4f$, Ce $5d$ and Fe $3d$ states.
 The Fermi energy is taken as the energy zero.
 The first and third panels show the majority spin channel, and the
 second and forth panels, the minority spin channel.
 In order to enhance visibility, the
 magnitude of the Ce $5d$ DOS (dashed line)
 has been multiplied by a factor of 10.
 }
 \end{figure}

In Fig.~\ref{dos.fig}, the spin-resolved partial DOS
for the Ce $4f$, Ce $5d$ and Fe $3d$ states are shown.
The first and third panels show the majority spin channel for Ce and
Fe, respectively, and the second and fourth panels show the minority
spin channel.
Comparing the DOS for Ce and Fe, we see that
the Ce $4f$ and Fe $3d$ states have opposite
spin polarization. Furthermore, the Ce $5d$ band width is
seen to be much larger than that of the
Ce $4f$ and Fe $3d$ states, with the magnitude of the Ce $5d$ DOS
roughly an order of magnitude smaller than that of
the Ce $4f$ and Fe $3d$ states.

\begin{figure}
  \centering\epsfxsize=0.5\textwidth \epsffile{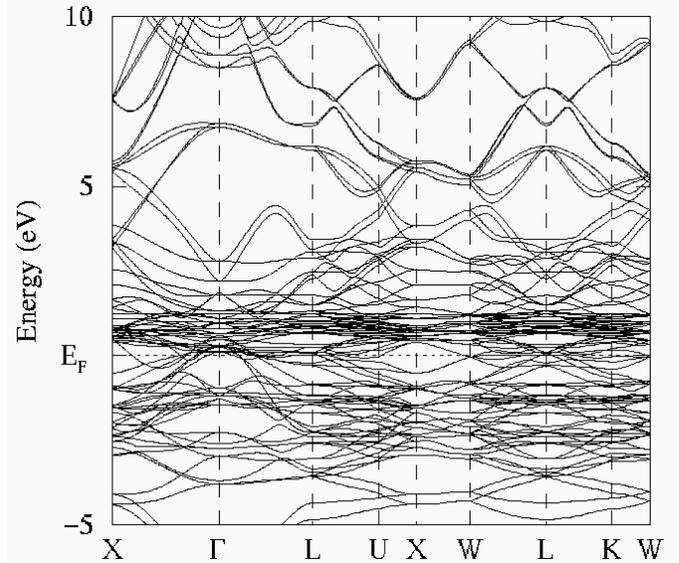}
 \caption
 {
 \label{band.fig}
 Energy bands for CeFe$_2$ along high-symmetry directions.
 The Fermi energy is taken as the energy zero.
 }
 \end{figure}

Fig.~\ref{band.fig} shows the band-structure of spin-polarized
CeFe$_2$ along
high-symmetry directions in the Brillouin zone.
The flat bands clustered just above the Fermi level are
predominantly of $4f$-character.
In the region from $\sim$2 eV to $\sim$10 eV,
and around -5 eV with respect to
the Fermi level, the spin splitting of the bands is clearly visible.

Our calculated total spin
magnetic moment amounts to 2.48 $\mu_{\rm B}$ per formula unit
(experimental saturation magnetization: 2.30 $\mu_{\rm B}$, as stated earlier
in this paper),
with the main contributions being the
Ce $4f$ moment -0.54  $\mu_{\rm B}$,
the Ce $5d$ moment -0.23  $\mu_{\rm B}$, and the
Fe $3d$ moment 1.75 $\mu_{\rm B}$.
The partial occupation numbers summed over spin,
within the muffin-tin spheres, are
1.07 for Ce 4$f$,
1.28 for Ce 5$d$,
and
6.18 for Fe 3$d$.
Note that the partial spin magnetic moments are calculated using the
partial occupation numbers inside the muffin-tin spheres,
which is somewhat arbitrary. The total spin moment is on the
contrary, of course, well defined.
An obvious point, which seems to have been overlooked so far, is that
not only in band-structure calculations, but
also experimentally, the division of space between
individual atomic species in a compound is in fact not unique
nor even well defined.
It is reasonable to believe that
different experimental procedures differ in their
``volume of sensitivity'' around each atom,
and thus effectively correspond
to different ways of dividing up the total space in the compound
between the atoms.
This could be one reason
why different experimental techniques
find quite different values for the partial magnetic moments,
and
also why, in order to find the total Ce moment from experimental results,
assumptions have to be made regarding the ratio of the
number of $5d$ and $4f$ electrons contributing to the magnetization.\cite{kennedy93}
%
%
%
An analysis of experimental data along this direction of thought
might help resolve
controversies regarding the electronic structure
of CeFe$_2$. \cite{delobbe99}


A calculation of the moments including spin-orbit coupling and
orbital polarization,\cite{op} using the FP-LMTO method,
has been performed earlier by Trygg {\it et al.}.\cite{trygg94}
The difference between the presently reported
spin moments and the ones reported by  Trygg {\it et al.},\cite{trygg94}
which include spin-orbit coupling,
is
very small, around 1\%. Thus, the effect of including spin-orbit
coupling is shown to have only a very minor effect on the magnitude
of the spin moment.
Calculations by Eriksson {\it et al.}\cite{eriksson88}
using the atomic sphere approximation (ASA)
give
somewhat different values for the magnetic moments than
the present method, in which no such geometrical approximation regarding the
form of the potential, wave functions or charge density is made.
As demonstrated in Ref.~\onlinecite{trygg94}, the spin density in
CeFe$_2$ is highly non-spherical, which may well be the reason for the
differences in results from full-potential and ASA calculations.
To summarize, the arguments presented above
justify our present calculational approach, i.e,
using a full-potential method, but neglecting spin-orbit coupling.


Concerning the absolute magnitudes of the individual moments,
the discrepancies between different experimental approaches can
be quite large, for instance, the Ce $4f$ spin moment is measured to
be -0.37 $\mu_{\rm B}$ with XMCD, whereas polarized neutrons find
the corresponding moment to be only about a fourth as large:
-0.10  $\mu_{\rm B}$.
As already touched upon above, one reason for these discrepancies
between different experimental techniques may well
be that they differ in the way
the space in the compound is effectively divided up between the atoms.
With this in mind, the magnitudes of our calculated moments must
be said to be in satisfactory agreement with experimental
findings, although the overall trend appears to be that
the calculations overestimate the moment magnitudes.

\begin{figure}
  \centering\epsfxsize=0.5\textwidth \epsffile{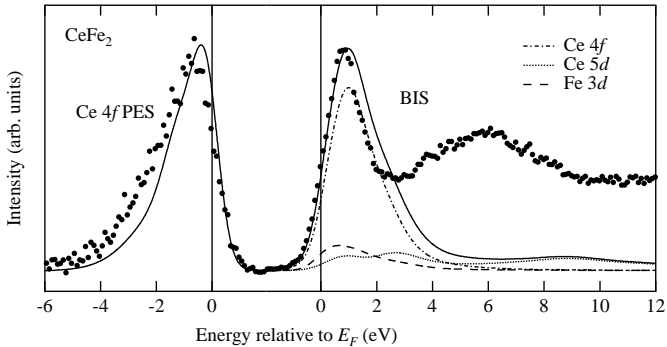}
\caption{
Comparison of the DFT DOS with the experimental spectra.
dot-dashed curves show orbital components. In the PES part, the Ce 4$f$
spectrum obtained by Ce 3$d$-4$f$ resonant PES is shown.
}
\label{dos_exp_4f.fig}
\end{figure}
In Fig.~\ref{dos_exp_4f.fig}, we compare the valence-band Ce 4$f$ PES and
BIS spectra with the DFT DOS. As mentioned earlier, the Ce 3$d$-4$f$
resonant PES spectrum and the BIS spectrum are
rather bulk sensitive, and thus it is relevant
to compare these spectra with bulk DFT calculations.
In the PES part of the spectra, comparison is
made between the experimentally obtained Ce 4$f$ spectra and the Ce
4$f$-projected DFT DOS. As for the BIS part, the Ce 4$f$, 5$d$ and Fe
3$d$ partial DOS have been added taking account of the atomic
photoionization cross-sections.~\cite{Yeh} Agreement between experiment
and theory is satisfactory almost to the same extent as in the SIAM
calculation.
In the DFT DOS, the structure around 6 eV in the BIS spectrum is of course not
reproduced, since this structure corresponds
to the 4$f^2$ final state, and thus is a purely excited-state property of
the system.
Furthermore, in the DFT calculation, the energy of the
near $E_F$ peak in the Ce 4$f$ PES spectrum and also in the BIS spectrum is
slightly higher than in the experimental spectra.
Also, the intensity on the higher
binding energy side of the PES spectrum is underestimated in the
calculation.
However, one should note that the relative intensities depend
also on the transition matrix elements, which are not included in the
DOS curves.

\begin{figure}
 \centering\epsfxsize=0.5\textwidth \epsffile{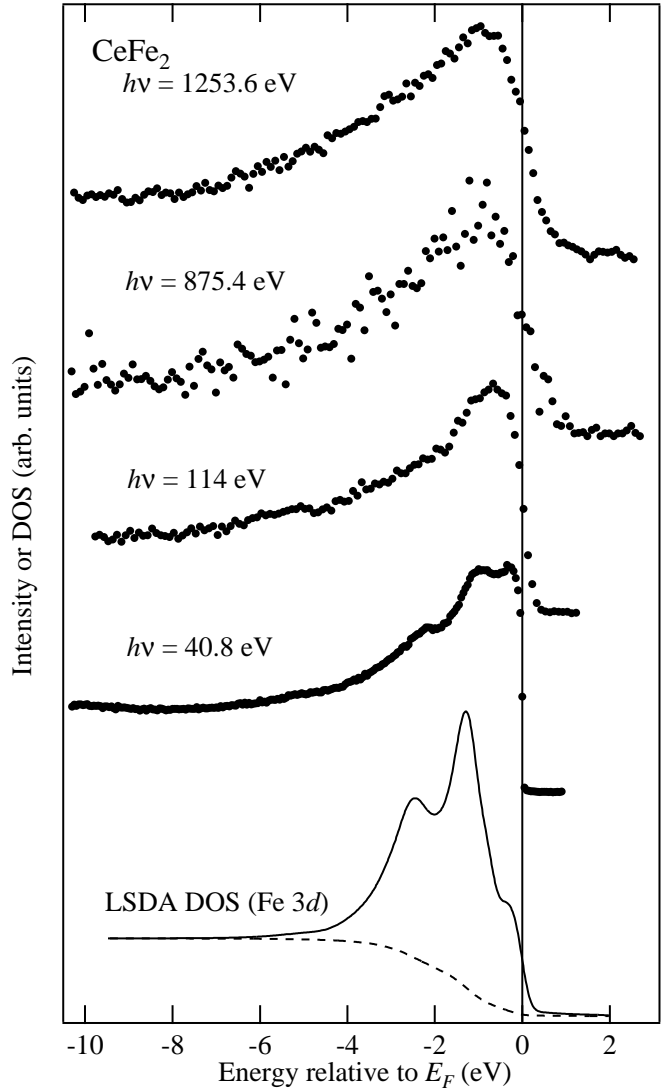}
\caption{
Valence-band PES spectra of CeFe$_2$ taken at various photon energies
compared with the Fe 3$d$ partial density of states from the band-structure
calculation. The DFT DOS
has been broadened with the experimental
resolution of the 114 eV spectrum.
}
\label{dos_exp_3d.fig}
\end{figure}

Fig.~\ref{dos_exp_3d.fig} shows the valence-band XPS, Ce 3$d$-4$f$ and 4$d$-4$f$
off-resonance and UPS (He~{\small II}) spectra. The spectral
weight comes primarily from the Fe 3$d$ states,
and thus we compare these spectra with the
Fe 3$d$ partial DFT DOS.
Although there are
differences in the surface sensitivity and in the contributions from
other orbitals, all the experimental spectra have similar band widths
and line shapes. In comparison with the DFT DOS, although overall features
are well reproduced, the experimental
spectra have larger spectral weight near $E_F$ than the calculated DOS.

\section{Discussion and Conclusions}
Generally, for any material, the surface electronic structure can
differ substantially from the bulk one.
For valence fluctuating systems like CeFe$_2$, it is very likely that
the electronic structure of the Ce atoms close to the surface is not the same
as that of the bulk Ce atoms.
Therefore, in order to study the bulk electronic
structure of valence fluctuating
Ce compounds by means of high-energy spectroscopic
methods, it is essential to take into account the effects of the sample
surface when interpreting the spectra.
In the present work, by assuming that the
spectra are superpositions of the surface and bulk components, we
have shown that all measured
spectra of CeFe$_2$ are fairly well reproduced by the SIAM calculations using the
same set of parameters. At the surface, Ce is found to be nearly
trivalent. The bulk set of parameters places CeFe$_2$ in the strongly
intermediate-valent regime, giving a $4f$-occupation number $n_f$ as
small as 0.78 (with considerable amplitude of the $f^2$ configuration
both in the surface and bulk). This means that the Ce 4$f$ states are
strongly hybridized with the Fe 3$d$ states in the bulk and that the
states around $E_F$ have a large amount of $f$ character.
Also, the differences between our SIAM-derived bulk- and surface spectra
for CeFe$_2$ are similar
to the differences between directly measured bulk- and surface spectra of
Ce-metal, a difference which can be explained as due to larger
$4f$ hybridization in the bulk than at the surface.

Apart from the SIAM analysis,
a number of features of the
measured spectra force us to draw the same conclusion
regarding the nature of the $4f$ states in the bulk and at the surface,
notably
the large difference between the Ce $4f$ spectra obtained
from the Ce $3d-4f$ and Ce $4d-4f$ resonant PES, and
the strong intensity of the peak near $E_F$ in the BIS spectrum,

As a result of the strong hybridization of the Ce 4$f$ states in the
bulk, the ``$f^0$'' final state feature in the Ce 4$f$ spectrum deduced
from the Ce 3$d$-4$f$ resonant PES has a very weak intensity in contrast
to the Ce 4$d$-4$f$ resonant PES spectrum, where the $f^0$-final-state
feature leads to the well-known double-peak structure. In such a case,
i.e, where there is strong screening of excitations,
it is expected that the bulk 4$f$ spectrum can be interpreted in terms of
a one-electron picture, and thus the DFT DOS should compare
well with the $4f$ spectrum. This is also seen to be the case.
Comparison of the 4$f$ spectra with the
DOS (Fig.~\ref{dos_exp_4f.fig}) shows that the Ce 4$f$ partial DFT DOS
describes the valence-band Ce 4$f$ spectra well, except for the $f^2$
structure in the BIS spectrum of course,
since the  $f^2$-peak is due to incomplete screening.
This fact poses the question of how the
``$f^1$'' final state (which is commonly referred to as the ``Kondo peak'')
of the SIAM picture and ``the 4$f$ band'' in the band picture are related
to each other,
since according to DFT, this peak is a one-electron feature, and
in the SIAM, this is due to a many-body effect.

We also draw the conclusion that
due to the strong hybridization in the bulk, the spin-orbit side band seen in
the high-resolution UPS spectra (Fig.~\ref{ups.fig}) must have surface
origin since it is known that the spectral weight of the spin-orbit side
band is strongly reduced when
the hybridization is strong.~\cite{Weschke2}

We now turn to a more detailed comparison of how
well the DFT calculations and the SIAM analysis perform
for the different spectra.
Regarding the position of the near $E_F$ peak in the BIS
spectrum, the DFT calculation predicts a higher energy than
experiment while the SIAM calculation predicts a lower energy than
experiment. The intensity on the higher binding energy side of the near
$E_F$ peak in the PES spectrum is underestimated in the DFT calculation
(although strictly speaking, intensities cannot be expected to be
reproduced with a DOS, since the transition matrix elements are neglected)
while it is overestimated in the SIAM calculation.
The position of the
near $E_F$ peak in the PES spectra is calculated to be too close to
$E_F$ both in the SIAM and DFT calculations.
As for the Fe 3$d$ component, the DOS calculated using DFT
does not give a good account
of the intensity in the experimental spectra near $E_F$.
This may be due to that the $4f$ transition matrix elements are
large close to the Fermi level compared to the $3d$ transition matrix elements.

All in all, the above discussion amounts to that the Ce $4f$ electrons
in the bulk hybridize strongly with the Fe $3d$ electrons. This
conclusion agrees perfectly with the experimentally observed
anti-parallel coupling of the Ce and Fe moments, which is also
reproduced in the DFT calculation.
In the SIAM, the $4f$ electron is assumed to be localized,
which indirectly implies that a parallel coupling of the Ce and Fe moments
is expected.

Finally, we wish to mention some sources of error in the present
work. Our measurements were done on scraped surfaces, which might
make the surface rough, and therefore ill-defined. Furthermore,
the precise values of  photoelectron mean free paths are
difficult to estimate, which naturally also has the effect of making the
border between ``bulk'' and ``surface'' somewhat ill-defined.
Our SIAM is not the most elaborate one, for instance we assume
a degeneracy of 14 of the $4f$ level, thereby neglecting spin-orbit
coupling and anisotropic hybridization effects, which leads to a crystal-field
splitting.
Furthermore, as in all DFT calculations,
the functional used treats electron
correlation only to a limited extent, i.e., it is not meaningful to
expect perfect agreement between the DFT results and
experiment.

\section*{Acknowledgment}
The authors thank the staff of SOR-RING, Y.~Azuma, T.~Miyahara and the
staff of Photon Factory for technical support. The authors also thank
A.~Sekiyama, J.~Okamoto, T.~Tsujioka and T. Saitoh for help in the experiment.
A.D. acknowledges financial support from the Swedish Foundation
for International
Cooperation in Research and Higher Education (STINT).
Part of this work has been done under the approval of the Photon Factory
Program Advisory Committee (Proposal No. 94G361).
A.D. and O.E are grateful to J.~M.~Wills for supplying the full potential code used in this
study.

\end{document}